\documentstyle[12pt,amssymb,amsfonts,amsbsy]{article}
\begin{document}
\sloppypar
\author{Ilja Schmelzer\thanks
       {WIAS Berlin}}

 \title{Realism And Empirical Evidence}

\begin{abstract}

We define realism using a slightly modified version of the EPR criterion of
reality.  This version is strong enough to show that relativity is
incomplete.

We show that this definition of realism is nonetheless compatible with the
general principles of causality and canonical quantum theory as well as with
experimental evidence in the (special and general) relativistic domain.

We show that the realistic theories we present here, compared with the
standard relativistic theories, have higher empirical content in the
strong sense defined by Popper's methodology.

\end{abstract}

\maketitle

\section{Introduction}

The violation of Bell's inequality \cite{Bell} predicted by quantum theory
shows an incompatibility between classical realism, causality and
relativistic quantum theory.  Thus, if we use a strong enough axiom system
for realism, we can prove that Einstein causality is false. In the first
section we give such a definition of realism, based on a minor modification
of the EPR criterion of reality.

Thus, we have a conflict between our definition of realism and relativity.
One very popular solution of this conflict is that this definition of
realism is too strong and should be weakened.  But there is also another
possibility: to prefer realism and to accept that relativity is incomplete.

It seems, the preference for the first solution is supported by a lot of
very different arguments as well as esthetic preferences, not by objective
comparison criteria.  In this paper, we apply Popper's scientific
methodology \cite{Popper} to compare above variants.  The advantage of this
methodology is that we do not have to rely on uncertain notions like
simplicity, beauty and so on, but have certain, well-defined criteria:
empirical falsification and empirical content.

Thus, at first we establish that our strong notion of realism is not
in contradiction with experiment.  This is done by the explicit
presentation of realistic theories for all important domains.

In the first step we present realistic Galilean invariant theories for the
relativistic domain.  This is the well-known Lorentz ether theory in the
domain of special relativity and a generalization named post-relativistic
gravity in the domain of strong gravitational fields.

What remains is the canonical quantization of Galilean invariant theories.
But for classical canonical quantum theories we have a realistic hidden
variable theory --- Bohmian mechanics.  Thus, the compatibility of the
related quantum theories with realism is not problematic.

Once we have found that realism is compatible with all empirical evidence,
to show the advantage of realism in predictive power is simple.  For this
purpose is seems sufficient to look at the definition of realism we have
given and to estimate how often we use these axioms in everyday reasoning.
To reject one of the axioms means that all these considerations become
invalid.

By comparison between these realistic theories with their relativistic
competitors we find also some other places where the realistic theory makes
stronger predictions.  For example, non-trivial topology of space-time is
forbidden, the part behind the horizon of a black hole cannot be reached.
These differences in empirical content are not very essential, they are only
of theoretical importance. But it is remarkable that they all are in favour
of realism.  There are no experiments which allow to falsify relativity
without falsification of these realistic theories too.  Thus, the comparison
of empirical content shows a clear advantage of realism.

\section{The Definition Of Realism}

For the purpose of this paper, we give a definition of realism based on a
minor but essential modification of the EPR criterion of reality. We specify
that the disturbance of the state mentioned in the EPR criterion should be a
real disturbance, thus, also an ``element of reality''. As a consequence,
the EPR criterion allows to prove the existence of some ``element of
reality'' --- if not a hidden variable, than a hidden causal disturbance.

The modified criterion is strong enough to falsify Einstein causality and to
prove the incompleteness of special relativity based on the violation of
Bell's inequality.  Note that this is not a non-trivial conclusion, but
simply a reformulation of the well-known incompatibility of local realism
with the violation of Bell's inequality. The possibility to define realism
in such a way is de-facto a tautology --- all we have to do is to close all
``loopholes'' introducing new axioms.  Thus, the only interesting point of
this section is the natural and simple character of our set of axioms and
their agreement with the common sense notion of realism.

\subsection{The EPR Criterion Of Reality}

First, let's reformulate the well-known EPR criterion of reality.  The
original formulation is the following \cite{EPR}:

{\it "If, without in any way disturbing a system, we can predict with
certainty ... the value of a physical quantity, than there exists an
element of physical reality corresponding to this physical quantity."}

The central objection in Bohr's reply was that the EPR reality criterion
``contains an ambiguity as regards the meaning of the expression `without in
any way disturbing a system'. Of course, there is ... no question of a
mechanical disturbance ... But ... there is essentially the question of an
influence on the very conditions which ... constitute an inherent element of
the description of any phenomenon to which the term `physical reality' can be
properly attached ...'' \cite{Bohr}.  

Our objection is very close.  We also feel some ambiguity in the expression
``in any way disturbing a system''.  We are interested in a description of
this disturbance in terms of realism.  The simple solution is that this
disturbance is also an ``element of reality'', only of a slightly different
type --- not an object, but a causal relation.

Thus, we have a measurement A which returns a value, we have a prediction
(also a measurement) B which also returns a value.  Now, let's introduce the
following denotations:

\begin{description}

\item[$A\equiv B$] denotes the observable fact of a 100
between the values returned by A and B.

\item[$\exists A\to B$] denotes the existence of a real disturbance of the
measurement B caused by A --- an ``element of reality''.  The phrase
``without in any way disturbing a system'' we translate as $\not\exists (B\to
A)$.  

\item[$\exists v(A)$] denotes the existence of an object, the
``element corresponding to'' the observed value.  This element is
predefined, independent of the choice of measurement A and B, that means,
independence axioms of classical probability theory may be applied.

\end{description}

Now note that B is named `prediction'', that means $t_B<t_A$.
Applying causality, we can conclude that A cannot disturb B too, thus,
$\not\exists (A\to B)$ too.  After this observation, the reference to
time ordering may be omitted.

Last not least, we put the two term $\not\exists A\to B$ and $\not\exists
B\to A$ on the other side.  We obtain the following reformulation of the
EPR criterion:

\begin{center}\it 
If $(A\equiv B)$ then $(\exists (A\to B)$ or $\exists (B\to A)$ or $\exists v(A))$.
\end{center}

\subsection{Completeness Of A Theory}

This formulation of the EPR criterion has a simple form: on the left side we
have the result of an observation. On the right side we have claims about
existence of some objects or relations.  We have three different
possibilities.  Nonetheless, in any case from the observation follows the
existence of some ``element of reality'' --- an object or a relation between
objects.  In other words, the EPR criterion tells that a {\it realistic
explanation} of every observable correlation exists.  This suggests a simple
and natural criterion of completeness of a realistic theory:

{\it A theory is complete if it gives a realistic explanation for every
observable correlation.}

The property of being a complete realistic theory has non-trivial empirical
content in Popper's sense.  Indeed, if the theory does not describe any of
the three alternative explanations, the theory predicts no correlation for
the related observation. Nonetheless, for a reasonable definition we have to
add some axioms about existing objects and relations:

{\it All axioms of classical logic and classical probability theory may be
applied to existing objects and relations.}

Indeed, without these properties it would not be justified to use the notion
``existence'' in the common sense of realism.

\subsection{Bell's Theorem}

With this definition, we have included all properties of the ``hidden
variables'' we need to prove Bell's inequality or one of it's various
variants:

{\it If $\exists v(A)$ then Bell's inequality is fulfilled.\/}

We do not consider here the details of this proof and the possibility
of loopholes of the existing experimental tests and assume that --- as
predicted by quantum theory --- Bell's inequality is violated in
reality even if the observations A and B are space-like separated.
Thus, it follows that $\not\exists v(A)$.  That means, the EPR
criterion proves that

{\it $\exists (A\to B)$ or $\exists (B\to A)$\/}

Relativity does not describe such causal relations.  It follows
immediately that Einstein causality is wrong.  Thus, if we want to
hold causality as a law of nature, we have to reject relativity.
Moreover, even if we reject causality, relativity is not a complete
realistic theory.  This immediately follows from our criterion of
completeness and the previous result.

Note that we have not used a theory of causality here --- all what we
have used from causality is the notion $A\to B$ for a causal relation
and the concept that such a causal relation is an element of reality.

\section{Compatibility Of Realism With Empirical Evidence}

Now, let's show that realism is compatible with all available empirical
evidence. 

\subsection{Lorentz Ether Theory}

Once we accept realism and the violation of Bell's inequality and do not
believe into closed causal loops, we can de-facto derive Lorentz-Poincare
ether theory \cite{Poincare}, that means special relativity with a hidden
preferred frame - the rest frame of the Lorentz ether.

Indeed, let's assume that for almost all pairs of events A and B we have
{\it $\exists (A\to B)$ or $\exists (B\to A)$\/}.  Moreover, let's assume the
elementary properties of causality like transitivity and the absence of
closed causal loops.

Now, let's define absolute future and past in the following way: For an
arbitrary event B we test the violation of Bell's inequality. After this, we
know $\exists (A\to B)$ or $\exists (B\to A)$.  Because the closed causal loop
$A\to B\to A$ is forbidden, only one of the two can be true. Now, if $A\to
B$ then B is in the future of A, and if $B\to A$, then B is in the past of
A.

If there are no closed causal loops, Bell's inequality should be fulfilled
at least in the degenerated case --- contemporaneity.  But the uncertainty
of time measurement leads to some difference in absolute time. Thus, this
degenerated case does not leads to observable effects.

Note that we cannot measure absolute contemporaneity --- we have no
way to detect what is the correct choice --- but, if we accept
realism, we have a proof of it's existence.  Thus, a complete
realistic theory should define the behaviour of these hidden
variables.

In the case of special relativity, this choice is simple: we assume
that absolute time coincides with coordinate time of some preferred
inertial system.  We obtain a well-known classical theory --- Lorentz
ether theory.

\subsection{Theory Of Gravity}

A generalization of Lorentz ether theory to gravity named post-relativistic
gravity has been defined by the author \cite{Schmelzer1},
\cite{Schmelzer2},
\cite{Schmelzer3}.
The theory is a Galilean-invariant ether theory, with an ether described by
positive density $\rho(x,t)$, velocity $v^i(x,t)$ and a positive-definite
stress tensor $\sigma^{ij}(x,t)$.  Interaction with the ether causes an
universal time dilation defined by the following Lorentz metric:

\[ g^{00}\sqrt{-g} = \rho \]
\[ g^{0i}\sqrt{-g} = \rho v^i \]
\[ g^{ij}\sqrt{-g} = \rho v^i v^j - \sigma^{ij}\]

The equations of this ether theory are the classical Einstein equations and
the harmonic coordinate equation:

\[ \partial_i (g^{ij} \sqrt{-g}) = 0 \]

which defines the classical conservation laws for the ether.  Thus, the
ether is no longer stationary, but is influenced by the matter.  This solves
a conceptual problem of Lorentz ether theory.  It also allows to define
local energy and momentum densities for the gravitational field, thus,
solves a problem of general relativity.

This theory coincides in almost all predictions with general relativity.  A
solution of post-relativistic gravity defines a solution of general
relativity.  This solution defines all classical observables of the
post-relativistic solution.  Thus, in the classical domain the empirical
content of general relativity is not greater: An observation which cannot be
described by general relativity cannot be described by post-relativistic
gravity too.

On the other hand, there are some interesting additional predictions of
post-relativistic gravity.  Solutions with non-trivial topology are
forbidden. Moreover, the notion of completeness is different. There is no
reason to assume that the ``metric'' defined by the ether should be
complete.  It may be incomplete, and in interesting cases like the collapse
into a black hole it is really incomplete as a solution of general
relativity: the complete post-relativistic collapse solution does not
contain the part behind the horizon --- the collapse stops in absolute time
immediately before horizon formation.  

Thus, the generalization of Lorentz ether theory to gravity is possible,
leads to a Galilean invariant ether theory which is in agreement with
experiment as well as general relativity.  Compared with general relativity
it has more empirical content.

This theory is obviously compatible with classical realism and classical
causality as well as Lorentz ether theory.

\subsection{Canonical Quantum Theory}

Because we have found classical realistic Galilean-invariant theories for
the relativistic domain too, we only have to consider the quantization of
classical Galilean invariant theories.  Thus, the canonical quantization
scheme may be used.

But in this case, the compatibility with realism is not problematic. To
prove this, it is sufficient to remember about the existence of Bohmian
mechanics \cite{Bohm}.  This is a deterministic hidden variable theory for
classical quantum theory.  It is obviously compatible with realism.

Thus, realism becomes problematic during quantization not because of
compatibility problems with quantum theory, but because
incompatibility with relativity.

This remains valid for the case of generalized Hamiltonian systems.
Indeed, they are normal Hamilton systems on a subspace, and that we
prefer to choose other coordinates is our choice, not a conceptual
difference between the theories.  That means, this generalization does
have any influence on the question we consider here --- compatibility
with realism.

An infinite number of steps of freedom is of course a technical problem,
but also orthogonal to our question.  The general problem of ultraviolet
infinities in field theories we discuss below for quantum gravity.

\subsection{Quantum Gravity}

The quantization of post-relativistic gravity does not have the conceptual
problems of the quantization of general relativity, like the problem of time
\cite{Isham}, topological foam, information loss for black holes, absence of
local energy and momentum densities for the gravitational field.

On the other hand, it leads to ultraviolet problems as well as general
relativity.  From point of view of the ether concept, this can be
easily explained with an atomic structure of the ether.  But this is
not the place to speculate about atomic ether models.

Indeed, we are interested here in compatibility questions. What we
want to show is the compatibility of empirical evidence and realism.
For this purpose we can use some very easy Galilean-invariant
regularization of post-relativistic gravity, for example a simple
regular lattice regularization.  To quantize this theory, we can apply
standard canonical quantization \cite{Schmelzer3}.  The resulting
theory, beautiful or not, is compatible with empirical evidence. 
It is also compatible with realism. 

\section{Discussion} 

Thus, we have shown the compatibility of realism with all available
empirical evidence.  We have proven this by the presentation of realistic
causal Galilean-invariant theories for all domains up to quantum gravity.

The empirical content of the presented theories is at least equal than
the empirical content of their relativistic competitors.  We have
found some non-trivial additional predictions. Thus, following
Popper's theory, we have to prefer realism as the theory with more
predictive power.

It can be said that the additional predictions are only of theoretical
importance. Indeed, the advantages of post-relativistic gravity become
obvious only in the quantum domain, but quantum gravity effects are
far away from our experimental possibilities.

But to save relativity we have to weaken our definition of realism.
This leads obviously to a very serious and important loss of empirical
content: we use all our axioms of realism in everyday reasoning.  If
we reject on of these axioms, we have to reject any argumentation
which uses this axiom too.

There is no necessity to consider the various different possibilities to
weaken some of our axioms of realism and causality to show that each of
these possibilities leads to less predictive power.  We can simply use the
fact that Bell's inequality is no longer valid in any of these weaker
theories to show that the predictive power of the weaker notion of realism
is really a weaker theory in the sense of Popper's criterion of empirical
content.

Indeed, if telepathy is possible, this will be in contradiction with
realism.  Telepathic effects may be observed by establishing
correlations which cannot be explained without such telepathic
possibilities.  But a special case of such a correlation is a
correlation which violates Bell's inequality.  Thus, if some
correlations between claims of two persons in different, isolated
rooms show a pattern which violates Bell's inequality, the realistic
theory is falsified.  But the theory with the weaker notion of realism
is not falsified, because Bell's inequality cannot be proven in this
theory.  This proves that the predictive power of realism is greater.

\end{document}